\documentclass{article}
\usepackage{arxiv}

\usepackage{geometry}
\usepackage{xcolor}
\usepackage[hidelinks]{hyperref}
\usepackage{booktabs}
\usepackage{graphicx}
\usepackage{amsmath}
\usepackage{amssymb}
\usepackage{bm}
\usepackage{array}
\usepackage{multirow}
\usepackage{caption}
\usepackage{url}
\usepackage{enumitem}
\usepackage{subcaption}
\usepackage{algorithm}
\usepackage{algorithmic}
\usepackage{lineno}
\newcommand{\Comment}[1]{\hfill$\triangleright$~#1}
\makeatletter
\let\ftype@table\ftype@figure
\makeatother

\title{SPINONet: Scalable Spiking Physics-informed Neural Operator for Computational Mechanics Applications}
\author{
  Shailesh Garg  \\
  Department of Applied Mechanics\\
  Indian Institute of Technology Delhi\\
  Hauz Khas, New Delhi 110016, India. \\
  \texttt{shaileshgarg96@gmail.com} \\
  \And
  Luis Mandl\\
  Institute of Structural Mechanics and Dynamics in Aerospace Engineering\\ Faculty of Aerospace Engineering and Geodesy\\
  University of Stuttgart, Stuttgart, Germany.\\\\
  Experimental Hepatobiliary Surgery Group\\
  Department of Hepatobiliary Surgery and Visceral Transplantation\\
  University of Leipzig Medical Center, Leipzig, Germany.\\\\
  Department of Civil and Systems Engineering\\ Johns Hopkins University, Maryland 21218, USA.\\
  \texttt{luis.mandl@isd.uni-stuttgart.de}\\
  \And
  Somdatta Goswami\\
  Department of Civil and Systems Engineering\\
  Data Science and AI Institute\\
  Johns Hopkins University, Maryland 21218, USA.\\
  \texttt{somdatta@jhu.edu}\\
  \And
  Souvik Chakraborty  \\
  Department of Applied Mechanics\\
  Yardi School of Artificial Intelligence (YScAI)\\
  Indian Institute of Technology Delhi\\
  Hauz Khas, New Delhi 110016, India. \\
  \texttt{souvik@am.iitd.ac.in}}
  
\begin{document}

\maketitle

\begin{abstract}
Energy efficiency remains a critical challenge in deploying physics-informed operator learning models for computational mechanics and scientific computing, particularly in power-constrained settings such as edge and embedded devices, where repeated operator evaluations in dense networks incur substantial computational and energy costs. To address this challenge, we introduce the Separable Physics-informed Neuroscience-inspired Operator Network (SPINONet), a neuroscience-inspired framework that reduces redundant computation across repeated evaluations while remaining compatible with physics-informed training. SPINONet incorporates regression-friendly neuroscience-inspired spiking neurons through an architecture-aware design that enables sparse, event-driven computation, improving energy efficiency while preserving the continuous, coordinate-differentiable pathways required for computing spatio-temporal derivatives. We evaluate SPINONet on a range of partial differential equations representative of computational mechanics problems, with spatial, temporal, and parametric dependencies in both time-dependent and steady-state settings, and demonstrate predictive performance comparable to conventional physics-informed operator learning approaches despite the induced sparse communication. In addition, limited data supervision in a hybrid setup is shown to improve performance in challenging regimes where purely physics-informed training may converge to spurious solutions. Finally, we provide an analytical discussion linking architectural components and design choices of SPINONet to reductions in computational load and energy consumption.
\end{abstract}
\section{Introduction}
Physics-informed deep learning \cite{raissi2019physics,karniadakis2021physics,raissi2018hidden,cho2023separable}offers an effective framework in
mechanical science, where observational data are limited, but the governing physical laws are well understood. By embedding these laws into the learning process, physics-informed deep learning models \cite{raissi2019physics,chakraborty2021transfer, goswami2020transfer,cho2023separable} restrict the space of admissible solutions and enable discretization-invariant surrogate modeling once training is complete, a desirable property in computational mechanics workflows. Incorporating operator learning \cite{lu2021learning,li2020fourier,tripura2023wavelet,kovachki2023neural} with physics-informed learning extends this paradigm by learning mappings between function spaces rather than individual solution instances, enabling rapid evaluation of solution fields for unseen but in-distribution inputs without re-training or re-solving the underlying PDE. This capability is especially valuable in applications such as parametric PDE analysis \cite{hesthaven2016certified,prud2002mathematical}, uncertainty quantification \cite{zhu2018bayesian,smith2024uncertainty}, and digital twin modeling \cite{rasheed2020digital,tao2018digital} in computational mechanics, where the input function space must be queried repeatedly across multiple evaluations.

Physics-informed operator-learning frameworks based on Deep Operator Networks (DeepONet)~\cite{mandl2025separable,wang2021learning,jiao2024solving}, Fourier Neural Operators \cite{li2024physics,eshaghi2025variational}, and Wavelet Neural Operators \cite{navaneeth2024physics,garg2025event} have demonstrated strong performance in learning various PDEs
Recent work has additionally focused on improving the scalability of these approaches for tackling high-dimensional problems. Within this context, the separable physics-informed deep operator network \cite{mandl2025separable} framework has gained attention. The framework leverages the idea of solving PDEs with separation of variables, thereby exploiting coordinate-wise factorization and enabling efficient residual computation via forward-mode automatic differentiation to reduce the cost of evaluating the PDE residual. 
However, in its vanilla form, these architectures rely on continuously active neurons, i.e., neurons are evaluated, and a full set of multiply-accumulate operations is executed at every forward evaluation, irrespective of the effective information content of the input. This results in increased computational load and hinders deployment under constrained computational or power budgets, as encountered in edge computing \cite{shi2016edge,wang2025empowering} or embedded \cite{an2023survey,ajani2021overview} environments.

To address this execution bottleneck in physics-informed learning, we focus on how information is represented and propagated, rather than modifying the operator’s functional form or the physics-informed loss. Sparse, event-driven neuron models~\cite{yamazaki2022spiking,nguyen2021review,dora2021spiking,putra2020fspinn}, inspired by biological computation, offer a principled mechanism to reduce redundant communication by activating neurons only when informative signals are present. However, their discontinuous dynamics pose challenges for residual-based physics-informed training, where accurate spatial and temporal derivatives are essential for enforcing physical constraints~\cite{neftci2019surrogate,garg2025event}.

Recent efforts have explored neuroscience-inspired physics-informed neural networks~\cite{wei2025physics,wang2023physics,tandale2024physics}, demonstrating that spiking or event-driven mechanisms can be incorporated into PINN-style frameworks. However, these approaches operate at the level of solution networks and do not address neural operators, which learn mappings between function spaces and are critical for parametric PDEs and multi-query evaluation settings. 
The only prior work integrating spiking dynamics with neural operators is the neuroscience-inspired WNO proposed in~\cite{garg2025event}, built upon the WNO backbone~\cite{tripura2023wavelet}. While it yields reasonably accurate result, its wavelet-based architecture differs structurally from DeepONet-style operators, which offer additional flexibility. Separable DeepONet variants, in particular, can seamlessly handle 
high-dimensional problems.
These considerations motivate the investigation of neuroscience-inspired DeepONet architectures, especially separable variants, as a structurally principled direction for enabling sparse, event-driven operator learning that remains fully compatible with residual-based physics-informed training and can scale to high dimensional problems.

In this paper, we propose the \emph{Separable Physics-informed and Neuroscience-inspired Operator Network (SPINONet)}. The central design principle of SPINONet is to resolve the apparent incompatibility between spiking neural dynamics and physics-informed operator learning through architectural separation. 
In separable physics-informed deep operator networks, physics-informed residuals require coordinate derivatives only through the coordinate-dependent components (trunk), while the input-function encoding (branch) enters the operator algebraically and remains independent of spatial and temporal differentiation. SPINONet leverages this structural decoupling to introduce spiking dynamics exclusively within the input-function encoding pathway, while preserving continuous, fully differentiable coordinate pathways required for physics-informed training.
Realizing sparse, event-driven computation in this setting, however, necessitates a spiking neuron model compatible with regression tasks and continuous function approximation. The class of such neuron models that can be trained natively and perform reliably in regression settings is limited, with primary candidates being the Variable Spiking Neuron (VSN)~\cite{garg2023neuroscience} and the Quadratic Integrate-and-Fire (QIF) model~\cite{wan2025lif}. While the proposed framework is compatible with either (and with future regression-friendly spiking models), we adopt VSN~\cite{garg2023neuroscience}, which supports graded, continuous-valued spikes and has demonstrated strong performance in regression and operator-learning contexts~\cite{jain2025hybrid,garg2024neuroscience,garg2023neuroscience}.
The key contributions of this work can be summarized as follows,
\begin{itemize}
\item \textbf{Architectural separation enabling physics-informed spiking operator learning.} 
We introduce \emph{SPINONet}, a separable physics-informed operator-learning framework that resolves the incompatibility between spiking dynamics and physics-based differentiation through structural decoupling. By confining spiking computation to the input-function encoding pathway, where inputs enter algebraically, and preserving continuous coordinate-dependent components for derivative evaluation, SPINONet enables sparse, event-driven computation without compromising operator expressivity or physics-informed training consistency. The coordinate-wise factorization further eliminates explicit full-mesh evaluations, ensuring smoothly scaling computational cost with increasing resolution.
\item \textbf{Accuracy retention with enhanced stability and scalability.} 
Across a diverse suite of time-dependent and steady-state PDEs, including high-dimensional spatio-temporal-parametric settings, SPINONet achieves predictive accuracy comparable to dense physics-informed operator-learning baselines while substantially reducing computational burden. Owing to its separable architecture, the framework scales smoothly with increasing dimension without incurring full mesh-based evaluation costs. We further demonstrate that incorporating a small amount of supervised data effectively mitigates degenerate solutions encountered in purely physics-informed regimes, improving stability and predictive fidelity without modifying the underlying physics-informed loss formulation.
\item \textbf{Analytical foundations for computational and energy efficiency.} 
We present a hardware-agnostic analytical framework that rigorously links sparse spiking activity and separable trunk evaluation to reductions in computational and energy cost. In addition, we formally analyze how forward-mode automatic differentiation synergizes with the proposed architecture to yield further efficiency gains, providing theoretical grounding for scalable operator learning.
\end{itemize}

The remainder of the paper is organized as follows. Section~\ref{section:background} reviews physics-informed operator learning and neuroscience-inspired computation. Section~\ref{section:methodology} introduces the SPINONet framework in detail. Section~\ref{section:results} presents and discusses the numerical experiments. Section~\ref{section:conclusion} concludes the paper and outlines directions for future work.

\section{Problem Statement}\label{section:background}
We consider parametric partial differential equations (PDEs) on a spatial domain $\Omega \subset \mathbb{R}^{\hat d}$ and, for time-dependent problems, a temporal domain $\mathcal{T} \subset \mathbb{R}$,
\begin{equation}
\begin{gathered}
\mathcal{N}\!\left(u(\bm x, t); \bm{\lambda}\right) = 0, 
\quad (\bm x, t) \in \Omega \times \mathcal{T},\quad
\mathcal{N}_{B}\!\left(u(\bm x, t); \bm{\lambda}\right) = 0,  
\quad (\bm x, t) \in \partial\Omega \times \mathcal{T}, \\
\mathcal{N}_{I}\!\left(u(\bm x, t); \bm{\lambda}\right) = 0, 
\quad (\bm x, t) \in \Omega \times \{t_0\}.
\end{gathered}
\label{eq:pde_general}
\end{equation}
where $\mathcal{N}$ denotes the governing differential operator, $\mathcal{N}_B$ and $\mathcal{N}_I$ encode boundary and initial conditions, and $\bm{\lambda}$ collects physical parameters. For steady-state problems, $\mathcal{T}$ and $\mathcal{N}_I$ are omitted.
Many computational mechanics and scientific computing applications require repeatedly solving such PDEs for varying inputs. This motivates learning the solution operator $\mathcal{G}:\mathcal{U}\rightarrow\mathcal{Y}$ that maps admissible input functions to solution fields. Physics-informed operator learning seeks a parametric surrogate $\mathcal{G}_{\boldsymbol{\theta}}\approx\mathcal{G}$ that generalizes to unseen inputs without re-solving the PDE.
DeepONets approximate nonlinear operators by expressing the solution at a location $(\bm x,t)$ as,
\begin{equation}
\mathcal{G}_{\theta}(\bm u)(\bm x,t)
=
\sum_{k=1}^{p} b_k(\bm u)\,\tilde t_k(\bm x,t),
\label{eq:deeponet}
\end{equation}
where the branch network, $b_k(\cdot)$ encodes the input $\bm u$ and the trunk network, $\tilde t_k(\bm x,t)$ encodes the spatial and temporal coordinates. 
In physics-informed DeepONet, the PDE is enforced via residual minimization, requiring spatial and temporal derivatives of $\mathcal{G}_{\theta}$.
%
%
While the vanilla physics-informed DeepONet performs well for a wide range of computational mechanics tasks, it is prohibitively expensive on large spatio-temporal grids. The coordinate-dependent network must be evaluated at every grid point, causing the cost to scale with grid resolution. Additionally, continuously activated neurons require full multiply-accumulate operations at each evaluation, regardless of input complexity, leading to high energy consumption during training and repeated inference.
The objective of this paper is to address these limitations by developing scalable operator-learning frameworks that reduce redundant neural activity and execution cost without modifying the operator representation or the formulation of the physics-informed loss.
\section{Proposed Framework}\label{section:methodology}
%
%
This section formally introduces the Separable Physics-informed Neuroscience-inspired Operator Network (SPINONet), a unified operator-learning framework that enables sparse, event-driven neural computation while preserving the mathematical structure required for physics-informed training. To substantiate the efficiency and scalability of SPINONet, we present three analytical studies: (i) a hardware-agnostic analysis of the execution benefits of sparse spiking activity, (ii) a quantitative assessment of the computational advantages of separable trunk evaluation, and (iii) a formal justification for employing forward-mode automatic differentiation in residual computation. Finally, we address the challenges posed by discontinuous spiking dynamics and describe how gradient-based training is achieved within the proposed architecture.
\subsection{Separable Physics-informed Neuroscience-inspired Operator Network}
%
We start by presenting the mathematical formulation of SPINONet, highlighting its separable construction, event-driven coefficient generation, and structural compatibility with residual-based physics enforcement.
Let $\mathcal{G}_\theta:\mathcal{U}\rightarrow\mathcal{Y}$ denote the solution operator mapping admissible inputs to the solution fields. For a $d$-dimensional spatio-temporal coordinate vector $\bm{\xi}\in\mathbb R^d$, vanilla separable DeepONet employs $d$ independent trunk networks, each taking a one-dimensional coordinate vector as input. Specifically, the $j$\textsuperscript{th} trunk network 
$\mathrm{tr}^{j}:\mathbb{R}\rightarrow \mathbb{R}^{pr}$, that takes $\xi_j\in \mathbb R$ as input and produces $pr$ basis functions, where $p$ denotes the latent dimension and $r$ the low-rank decomposition rank.  

For each coordinate direction $j\in\{1,\dots,d\}$, let $k_j\in\{1,\dots,n_j\}$ denote the grid index along the $\xi_j$ coordinate.
Evaluating $\bm{\mathrm{tr}}^j(\xi_j)$ at $n_j$ discrete locations $\hat{\bm \xi}_j = \{\xi_j^{(k_j)}\}_{k_j=1}^{n_j}$ along the $\xi_j$ axis yields an output $\bm{\hat{\mathrm{tr}}}^j(\hat{\bm\xi}_j)$ of size $n_j\times pr$, later reshaped to $n_j\times p\times r$. Given a set of one-dimensional coordinate grids, along each axis, the $d$ trunk networks therefore generate coordinate-wise embeddings that can be combined through a separable outer-product construction. The resulting elements of the basis functions at a location $\bm\xi = (\xi_1^{(k_1)},\dots,\xi_d^{(k_d)})$ of the spatio-temporal grid are defined as,
\begin{equation}
\tilde t_{k_1,\dots,k_d,m}(\bm \xi)
=
\sum_{i=1}^{r}
\left(
\prod_{j=1}^{d}
\hat{\mathrm{tr}}^{j}_{k_j,m,i}(\hat{\bm \xi}_j)
\right),
\quad m=1,\dots,p.
\end{equation}
Using the elements computed in the above equation, the final trunk output $\widetilde{\mathcal T}\in\mathbb R^{n_1\times n_2\times \dots\times n_d\times p}$ is constructed over the full lattice grid, despite each trunk network being evaluated only on one-dimensional coordinate samples along a single axis.
Given an input function $\bm u$, the branch network $\mathcal {B}(\bm u)\in\mathbb{R}^{p}$ outputs $p$ coefficients; and now, the operator output at a particular location is obtained by contracting the branch and trunk representations along the latent dimension $p$ as, 
\begin{equation}
\mathcal{G}_{\theta}(\bm u)(\bm{\xi})
=
\sum_{m=1}^{p}\mathcal B_m(\bm u)\,\tilde t_{k_1,\dots,k_d,m}(\bm \xi).
\label{eq:spinet_operator}
\end{equation}
The solution tensor of size $ n_1\times\cdots\times n_d$ over the whole spatio-temporal domain $\Omega\times\mathcal T$, can be obtained as,
\begin{equation}
\mathcal{G}_{\theta}(\bm u)
=
\sum_{m=1}^{p}\mathcal B_m(\bm u)\,
\widetilde{\mathcal T}_{\dots,m}.
\end{equation}
Physics-informed training is achieved by enforcing the governing equations in a residual sense. Given the operator approximation in Eq. \eqref{eq:spinet_operator}, the interior residual at a collocation point $\bm{c}\in\Omega\times\mathcal{T}$ is defined as
\begin{equation}
\mathcal{R}_{\theta}(\bm{c};\bm u)
=
\mathcal{N}\!\left(\mathcal{G}_{\theta}(\bm u)(\bm{c});\bm{\lambda}\right),
\end{equation}
and the corresponding interior loss is given by
\begin{equation}
\mathcal{L}_{\mathrm{int}}(\theta)
=
\frac{1}{N_c}
\sum_{i=1}^{N_c}
\left\|
\mathcal{R}_{\theta}(\bm{c}_i;\bm u)
\right\|^2.
\end{equation}
Boundary and initial condition constraints are enforced analogously by defining residuals associated with the boundary operator $\mathcal{N}_B(\cdot)$ and the initial condition operator $\mathcal{N}_I(\cdot)$ at collocation points on $\partial\Omega\times\mathcal{T}$ and $\Omega\times\{t_0\}$, yielding losses $\mathcal{L}_{\mathrm{bc}}(\theta)$ and $\mathcal{L}_{\mathrm{ic}}(\theta)$. The total physics-informed loss is then
\begin{equation}
\mathcal{L}_{\mathrm{phys}}(\theta)
=
\mathcal{L}_{\mathrm{int}}(\theta)
+
\lambda_{\mathrm{bc}}\,\mathcal{L}_{\mathrm{bc}}(\theta)
+
\lambda_{\mathrm{ic}}\,\mathcal{L}_{\mathrm{ic}}(\theta),
\end{equation}
where $\lambda_{\mathrm{bc}}$ and $\lambda_{\mathrm{ic}}$ balance the enforcement of boundary and initial conditions relative to the interior residual.
A defining structural property of the operator representation \eqref{eq:spinet_operator} is that differentiation with respect to the spatio-temporal coordinates acts exclusively on the coordinate-dependent trunk components. Specifically,
\begin{equation}
\nabla_{\bm{\xi}} \mathcal{G}_{\theta}(\bm u)(\bm{\xi})
=
\sum_{m=1}^{p}
\mathcal B_{m}(\bm u)\,
\nabla_{\bm{\xi}}
\tilde t_{k_1,\dots,k_d,m}(\bm{\xi}).
\label{eq:spinet_grad}
\end{equation}
The branch coefficients $\mathcal B_{m}(\bm u)$ remain independent of the spatio-temporal coordinates $\bm{\xi}$. This structural decoupling enables efficient evaluation of physics-informed residuals using forward-mode automatic differentiation, which is critical for scalable training in high-dimensional settings. Since coordinate derivatives do not act on the coefficient of the branch network, modifications to the input-function encoding pathway do not interfere with the spatio-temporal derivatives required for residual evaluation. SPINONet leverages this property to introduce sparsity-promoting VSNs at the level of coefficient generation while preserving physics-consistent training. In the proposed framework, the branch coefficients are generated using VSNs as the activation function, which support graded, continuous-valued information propagation while exhibiting sparse firing behavior. The dynamics of a VSN are governed by
\begin{equation}
{m}^{(\tau)}
=
\beta_l\,{m}^{(\tau-1)}
+
{z}^{(\tau)},
\end{equation}
where ${z}^{(\tau)}$ denotes the presynaptic input at spike time step\footnote{In spiking models, neuron dynamics are evaluated over a discrete sequence of spike time steps. We denote the total number of such steps (i.e., the spike-train length) by $T_s$, and index individual spike time steps by $\tau=1,\dots,T_s$. Unless stated otherwise, we use $T_s=1$, which yields sparse, event-driven computation while avoiding additional time-unrolling overhead during training and inference.} $\tau$, ${m}^{(\tau)}$ is the membrane potential, and $\beta_l\in(0,1)$ is a leakage parameter. The neuron produces an output
\begin{equation}
\begin{gathered}
\widetilde{{y}}^{(\tau)} = \mathbb H({m}^{(\tau)} - \mathcal{T}_h)\\
{y}^{(\tau)} = \phi({z}^{(\tau)}\widetilde{{y}}^{(\tau)})
\end{gathered}
\end{equation}
where ${y}^{(\tau)}$ is the neuron output at $\tau$ spike time step, $\mathbb H(\cdot)$ denotes the Heaviside function, $\mathcal{T}_h$ is a firing threshold and $\phi(\cdot)$ is a continuous activation function. In the event of a spiking event, i.e. $\widetilde y^{(\tau)} = 1$, the memory of VSN is reset to zero. The graded spiking outputs obtained after the last activation of the branch net are aggregated to form the coefficient vector $\mathcal B(\bm u)=[b_1(\bm u),\dots,b_p(\bm u)]^\top$, enabling sparse neural computation while maintaining suitability for regression-based operator learning.
%
%
Since VSN dynamics involve discontinuous spike-generation mechanisms, special care is required for gradient-based training; this issue is addressed in Section~\ref{subsec:surrogate}.

Training is primarily performed by minimizing the physics-informed objective $\mathcal{L}_{\mathrm{phys}}(\theta)$ defined above. However, in certain regimes, purely physics-informed training may converge to degenerate or trivial solutions. When a small amount of paired training data $\{(\bm u^{(i)},\bm y^{(i)})\}_{i=1}^{N_d}$ along with the grid where the target solution is obtained  is available, this behavior can be mitigated by augmenting the loss with an additional data supervision term,
\begin{equation}
\mathcal{L}_{\mathrm{data}}(\theta)
=
\frac{1}{N_d}
\sum_{i=1}^{N_d}
\left\|
\mathcal{G}_{\theta}(\bm u^{(i)}) - \bm y^{(i)}
\right\|^2.
\end{equation}
The resulting training objective is then given by
\begin{equation}
\mathcal{L}(\theta)
=
\mathcal{L}_{\mathrm{phys}}(\theta)
+
\lambda_{\mathrm{data}}\,\mathcal{L}_{\mathrm{data}}(\theta),
\end{equation}
where $\lambda_{\mathrm{data}}$ controls the contribution of the data term. In the absence of training data, we set $\lambda_{\mathrm{data}}=0$ and recover purely physics-informed training.
Algorithm~\ref{alg:spinet_training} summarizes the complete training procedure for SPINONet, including residual evaluation through the separable trunk structure and the generation of branch coefficients using VSN-based sparse computation.
\begin{algorithm}[ht!]
\caption{Training Algorithm for SPINONet}
\label{alg:spinet_training}
\begin{algorithmic}[1]
\REQUIRE
Training inputs $\{\bm u^{(i)}\}_{i=1}^{N_u}$;
interior collocation points $\{\bm{c}^{(j)}\}_{j=1}^{N_c}\subset\Omega\times\mathcal{T}$;
boundary collocation points $\{\bm{c}_b^{(j)}\}_{j=1}^{N_b}\subset\partial\Omega\times\mathcal{T}$;
initial collocation points $\{\bm{c}_0^{(j)}\}_{j=1}^{N_0}\subset\Omega\times\{t_0\}$;
(optional) paired data $\{(\bm u^{(i)},\bm y^{(i)})\}_{i=1}^{N_d}$ along with corresponding grid;
separation rank $r$; latent dimension $p$;
loss weights $\lambda_{\mathrm{bc}}, \lambda_{\mathrm{ic}}, \lambda_{\mathrm{data}}$.

\STATE Initialize trunk and branch network parameters.

\FOR{each training iteration}
    \STATE Sample a minibatch of inputs $\{\bm u^{(i)}\}$.
    
    \FOR{each input function $\bm u^{(i)}$ in the minibatch}
        \STATE Compute branch coefficients $\mathcal B(\bm u^{(i)})\in\mathbb{R}^{p}$ using the VSN-based branch network. In this step, the Heaviside function to compute $\widetilde y^{(\tau)}$ within a VSN is retained in its original form in this step.
        \STATE Evaluate coordinate-wise trunk networks to obtain trunk embeddings.
        \STATE Evaluate the operator output $\mathcal{G}_{\theta}(\bm u^{(i)})(\bm{\xi})$ on interior, boundary, and initial collocation points.
        \STATE Compute physics-informed residuals via forward-mode automatic differentiation. \\\Comment{Algorithm~\ref{alg:spinet_fad}}
    \ENDFOR
    
    \STATE Form physics-informed loss $\mathcal{L}_{\mathrm{phys}}=\mathcal{L}_{\mathrm{int}}+\lambda_{\mathrm{bc}}\mathcal{L}_{\mathrm{bc}}+\lambda_{\mathrm{ic}}\mathcal{L}_{\mathrm{ic}}$.
    \STATE If paired data are available, augment the loss with a data supervision term.
    \STATE Compute gradients of loss function w.r.t. trainable parameters of the SPINONet. For computing gradients through branch network, surrogate gradients must be used to approximate gradients of Heaviside function.\Comment{Algorithm~\ref{alg:spinet_surr}}
    \STATE Update all trainable parameters using gradient-based optimization.
\ENDFOR
\end{algorithmic}
\textbf{Outcome}: Trained SPINONet parameters $\theta$.
\end{algorithm}
\subsection{Energy Efficiency of Sparse Spiking Computation}
The primary motivation for using VSNs in SPINONet is the principle of sparse communication, widely regarded as a key mechanism for enabling energy-efficient information processing in biological neural systems \textit{in vivo}. By restricting neural activity to discrete spike events, spiking models reduce the number of active computations, thereby offering the potential for substantially lower energy consumption. Since absolute energy measurements depend on hardware architecture and implementation details, we adopt a conservative, hardware-agnostic analysis that isolates dominant contributors to energy expenditure. In particular, we focus on arithmetic operations and memory read/write activity, which dominate energy consumption on modern digital platforms.

We compare a densely connected artificial neural network (ANN) layer and a densely connected VSN layer with input and output dimensions, $N_{\text{in}}$ and $N_{\text{out}}$, respectively. For the VSN layer, computation unfolds over $T_s$ spike time steps, and input activity is assumed sparse. Total energy is decomposed into compute and memory components,
\begin{equation}
E_{\text{total}} = E_{\text{ops}} + E_{\text{mem}},
\end{equation}
where $E_{\text{ops}}$ represents the energy spent in synaptic operations and $E_{\text{mem}}$ represents the energy spent in memory related operations.
In a dense ANN layer, inference is executed as a matrix-vector multiplication, resulting in a fixed number of multiply-accumulate (MAC) operations,
\begin{equation}
\text{MAC}_{\text{ANN}} = N_{\text{in}}N_{\text{out}}.
\end{equation}
By contrast, the VSN layer performs synaptic computation only when spikes occur.
Let $\alpha_{\text{in}}$ denote the average input spiking activity per time step. The total number of input spikes over a window of $T_s$ spike time steps is modeled as
\begin{equation}
\theta_{l-1} = N_{\text{in}}T_s\alpha_{\text{in}}.
\end{equation}
Each spike fans out to all output neurons, yielding synaptic MACs equal to $\theta_{l-1}N_{\text{out}}$. In addition, each output neuron incurs one multiplication per spike time step due to leakage dynamics. Thus, the total MAC count for the VSN layer is
\begin{equation}
\text{MAC}_{\text{VSN}}
=
\theta_{l-1}N_{\text{out}} + T_sN_{\text{out}}.
\end{equation}
When $\alpha_{\text{in}}\ll 1$, synaptic computation scales with activity rather than full layer size, leading to substantially fewer multiplications than dense ANN execution.
To account for additional control and integration overhead, we include accumulation (ACC) operations. For ANN layers, this overhead is small and deterministic, while for VSN layers it scales with the number of time steps and spike-triggered updates. We model these costs as
\begin{equation}
\text{ACC}_{\text{ANN}} = N_{\text{out}} + (N_{\text{in}} + N_{\text{out}}),
\qquad
\text{ACC}_{\text{VSN}} =
2T_sN_{\text{out}} + \theta_{l-1}N_{\text{out}}.
\end{equation}
Although this introduces overhead absent in ANN layers, ACC operations typically consume significantly less energy than MACs and memory access operations, so reductions in MACs and memory traffic remain the dominant factor driving energy savings in sparse regimes.
Memory access energy is modeled by counting SRAM read and write operations. For ANN inference, inputs are read once, and all weights and biases are read once,
\begin{equation}
\text{Rd}_{\text{ANN}} = N_{\text{in}} + (N_{\text{in}}+1)N_{\text{out}},
\qquad
\text{WrOut}_{\text{ANN}} = N_{\text{out}}.
\end{equation}
For the VSN layer, input reads occur only when spikes arrive, and synaptic weight reads are similarly spike-triggered. Conservatively, we assume membrane potentials must be stored in SRAM and accessed at every spike time step, and that each neuron requires threshold and leakage parameters. The total read count is therefore
\begin{equation}
\text{Rd}_{\text{VSN}}
=
\theta_{l-1} + (\theta_{l-1}+1)N_{\text{out}} + T_sN_{\text{out}} + 2N_{\text{out}}.
\end{equation}
Energy is also expended on write operations. The VSN layer writes output spikes and updated neuron states, giving
\begin{equation}
\text{WrOut}_{\text{VSN}} = \theta_l + T_sN_{\text{out}},
\end{equation}
where $\theta_l$ denotes the total number of output spikes produced over the $T_s$ spike time steps.
Let $E_{\text{MAC}}$ and $E_{\text{ACC}}$ denote the energy cost of a single MAC and ACC operation, and let $E_{\text{Rd}}$ and $E_{\text{Wr}}$ denote the energy cost of a single SRAM read and write. Then, for $i\in\{\text{ANN},\text{VSN}\}$, the compute and memory energy are modeled as
\begin{equation}
E_{\text{ops}_i} = E_{\text{MAC}}\text{MAC}_i + E_{\text{ACC}}\text{ACC}_i,
\qquad
E_{\text{mem}_i} = E_{\text{Rd}}\text{Rd}_i + E_{\text{Wr}}\text{WrOut}_i,
\end{equation}
where $E_{\text{MAC}}$,  $E_{\text{ACC}}$,  $E_{\text{Rd}}$ and $E_{\text{Wr}}$ represent the energy spent in MAC, ACC, memory read and memory write operations respectively.
This model is intentionally pessimistic for VSN execution because it assumes frequent state read/write operations; in practice, architectures may retain neuron states in registers or local buffers, further reducing memory energy.
Using representative energy parameters of 45~nm CMOS technology \cite{jouppi2021ten}, analysis with varying spiking activity and number of nodes in layers shows that energy parity between dense ANN and VSN execution is approached only at high activity levels, approximately $\sim90\%$ for $T_s=1$ and $\sim86\%$ for $T_s=2$.

Overall, the discussion above indicates that cumulative spiking activity directly governs energy consumption, since both arithmetic operations and memory accesses scale with the total number of spikes. Consequently, total spiking activity provides a physically meaningful and practically useful proxy for energy efficiency. It is worth noting that model-level optimizations, such as pruning and quantization, can further reduce energy consumption in dense artificial neural networks, and these optimizations are also applicable to spiking neural networks \cite{li2022quantization,wei2025qp,jiang2025spatio,chen2021pruning}. Since such techniques are orthogonal to the execution paradigm, they are not explicitly considered here in order to isolate the energy implications of sparse, event-driven computation.

\subsection{Energy Efficiency of Separable Trunk Evaluation}
In this section, we discuss the energy implications of separable trunk evaluation. A complementary and orthogonal source of energy efficiency in SPINONet arises from the separable structure of the operator architecture, which substantially reduces the number of coordinate-dependent network evaluations required to produce a solution field. In a conventional, non-separable, DeepONet architecture, the trunk network is evaluated explicitly at every spatio-temporal query point on a discretized grid. For a $d$-dimensional grid with $N_1,\dots,N_d$ points along each coordinate axis, this results in $\prod_{j=1}^d N_j$ trunk evaluations, each incurring a full forward pass through the trunk network and associated arithmetic and memory costs.

By contrast, in separable operator architectures, coordinate-dependent embeddings are computed independently along each axis and combined algebraically through outer-product constructions. As a result, the total number of trunk network evaluations scales as $\sum_{j=1}^d N_j$, rather than $\prod_{j=1}^d N_j$. This reduction replaces an exponential dependence on dimensionality with a linear one, yielding orders-of-magnitude savings in the number of neural network evaluations for moderately sized grids.
To make this reduction more concrete, we consider the arithmetic cost of evaluating a single dense layer of the trunk network. Let the trunk layer consist of $N$ neurons with fully connected inputs of dimension $N_{\text{in}}$. A single evaluation of this layer incurs $N_{\text{in}}N$ multiply-accumulate (MAC) operations, along with $\mathcal{O}(N)$ accumulation operations.
In a conventional, non-separable DeepONet, this trunk layer is evaluated at every spatio-temporal grid point. For an $d$-dimensional grid with $N_1,\dots,N_d$ points along each coordinate axis, the total MAC count for a single trunk layer scales as
\begin{equation}
\text{MAC}_{\text{trunk}}^{\text{vanilla}}
=
N_{\text{in}}N
\prod_{j=1}^{d} N_j,
\end{equation}
with a corresponding accumulation cost that scales proportionally.
By contrast, in a separable operator architecture, the same trunk layer is evaluated independently along each coordinate axis. The resulting total MAC count scales as
\begin{equation}
\text{MAC}_{\text{trunk}}^{\text{separable}}
=
N_{\text{in}}N
\sum_{j=1}^{d} N_j,
\end{equation}
since coordinate-wise embeddings are computed once per axis and combined algebraically to form the full grid representation. Accumulation costs again follow the same scaling behavior. It is worth noting that the outer-product construction used to assemble the full grid representation introduces additional algebraic operations. However, these operations scale with the latent rank and the number of grid points only through simple tensor contractions, and do not involve additional neural network evaluations. As a result, their cost is negligible compared to the savings obtained by avoiding full trunk forward passes over the spatio-temporal mesh.

This comparison highlights that separable trunk evaluation replaces a multiplicative dependence on grid resolution with an additive one, yielding orders-of-magnitude reductions in arithmetic operations and reduced activation memory traffic in practice for moderately to large-sized spatio-temporal grids.
These reductions are independent of spiking activity and arise purely from the architectural structure of the operator representation.
%

\subsection{Forward-Mode Automatic Differentiation and SPINONet}
In previous sections, we analyzed the energy efficiency of SPINONet primarily from an inference perspective, focusing on arithmetic operations and memory access patterns induced by sparse spiking activity and separable trunk evaluation. Here, we explore the benefits of forward mode automatic differentiation (FAD) over reverse mode AD (RAD) under the condition of the input dimension being less than the output dimension. We also discuss how the separable trunk architecture of SPINONet amplifies this advantage compared to a non-separable architecture. 

To begin our discussion, let's first assume input $\bm u\in\mathbb R^{\mathfrak n_{in}}$ and output $\bm s\in\mathbb R^{\mathfrak n_{out}}$ and a network with $L$ hidden layers $ h_i\in\mathbb R^{\mathfrak n_i}, i = 1,\dots,L$. Now using chain rule, $\dfrac{\partial \bm s}{\partial \bm u}$ can be written as,
\begin{equation}
\frac{\partial \bm{s}}{\partial \bm{u}}
=
\left(\frac{\partial \bm{s}}{\partial \bm{h}_L}\right)_{\mathfrak n_{out} \times \mathfrak n_L}
\left(\frac{\partial \bm{h}_L}{\partial \bm{h}_{L-1}}\right)_{\mathfrak n_L \times \mathfrak n_{L-1}}
\cdots
\left(\frac{\partial \bm{h}_2}{\partial \bm{h}_1}\right)_{\mathfrak n_2 \times \mathfrak n_1}
\left(\frac{\partial \bm{h}_1}{\partial \bm{u}}\right)_{\mathfrak n_1 \times \mathfrak n_{in}}.
\end{equation}
The difference in FAD and RAD comes in the order in which the above Jacobian is computed. Visually, in FAD, the computation flow can be written as,
\begin{equation}
\frac{\partial \bm{s}}{\partial \bm{u}}
=
\frac{\partial \bm{s}}{\partial \bm{h}_L}
\Bigg(
\frac{\partial \bm{h}_L}{\partial \bm{h}_{L-1}}
\Big(
\cdots
\big(
\frac{\partial \bm{h}_3}{\partial \bm{h}_2}
\big(
\frac{\partial \bm{h}_2}{\partial \bm{h}_1}
\frac{\partial \bm{h}_1}{\partial \bm{u}}
\big)
\big)
\cdots
\Big)
\Bigg),
\end{equation}
and in RAD can be written as,
\begin{equation}
\frac{\partial \bm{s}}{\partial \bm{u}}
=
\Bigg(
\Big(
\cdots
\big(
\big(
\frac{\partial \bm{s}}{\partial \bm{h}_L}
\frac{\partial \bm{h}_L}{\partial \bm{h}_{L-1}}
\big)
\frac{\partial \bm{h}_{L-1}}{\partial \bm{h}_{L-2}}
\big)
\cdots
\Big)
\frac{\partial \bm{h}_2}{\partial \bm{h}_1}
\Bigg)
\frac{\partial \bm{h}_1}{\partial \bm{u}}.
\end{equation}
Thus, once the individual Jacobians are realized, the total computational cost of forming the final Jacobian can be expressed in terms of the number of multiply-accumulate (MAC) operations. For FAD, MAC operations $\mathrm{MAC}_{\mathrm{FAD}}$ can be computed as below,
\begin{equation}
\begin{gathered}
\mathrm{MAC}_{\mathrm{FAD}}
=
\mathfrak n_{out} \mathfrak n_L \mathfrak n_{in}
+ \mathfrak n_L \mathfrak n_{L-1} \mathfrak n_{in}
+ \cdots
+ \mathfrak n_3 \mathfrak n_2 \mathfrak n_{in}+ \mathfrak n_2 \mathfrak n_1 \mathfrak n_{in} \\
\mathrm{MAC}_{\mathrm{FAD}}
=
\mathfrak n_{out} \, \mathfrak n_L \, \mathfrak n_{in}
+
\sum_{\ell=2}^{L}
\mathfrak n_{\ell} \, \mathfrak n_{\ell-1} \, \mathfrak n_{in}.
\end{gathered}
\end{equation}
Similarly, MAC operations in RAD, $\mathrm{MAC}_{\mathrm{RAD}}$ can be computed as,
\begin{equation}
\begin{gathered}
\mathrm{MAC}_{\mathrm{RAD}}
=
\mathfrak n_{out} \mathfrak n_L \mathfrak n_{L-1}
+ \mathfrak n_{out} \mathfrak n_{L-1} \mathfrak n_{L-2}
+ \cdots
+ \mathfrak n_{out} \mathfrak n_1 \mathfrak n_{in} \\
\mathrm{MAC}_{\mathrm{RAD}}
=
\mathfrak n_{out} \, \mathfrak n_1 \, \mathfrak n_{in} + \sum_{\ell=2}^{L}
\mathfrak n_{out} \, \mathfrak n_{\ell} \, \mathfrak n_{\ell-1}.
\end{gathered}
\end{equation}
From the preceding expressions, the computational asymmetry between forward- and reverse-mode automatic differentiation becomes explicit, the leading-order cost of FAD scales linearly with the input dimension $\mathfrak n_{in}$, whereas the corresponding cost of reverse-mode AD scales linearly with the output dimension $\mathfrak n_{out}$\cite{baydin2018automatic,3119199}.
This distinction carries over directly to gradient evaluation via automatic differentiation, where FAD evaluates Jacobian-vector products, and reverse-mode AD evaluates vector-Jacobian products. These scaling behaviors are well established in the literature~\cite{cho2023separable,griewank2008evaluating,baydin2018automatic}, and imply that FAD is computationally and memory efficient when the final Jacobian is tall, i.e., $\mathfrak n_{in} \ll \mathfrak n_{out}$, while reverse-mode AD is more appropriate in the complementary regime, i.e., $\mathfrak n_{in} \gg \mathfrak n_{out}$.

We now discuss the benefit of forward-mode automatic differentiation formally when combined with SPINONet's separable trunk structure. The operator approximation from Eq. \eqref{eq:spinet_operator} admits a separable form, 
where the branch coefficients $\mathcal B_m(\bm u)$ depend only on the input function, and all coordinate dependence is confined to the trunk representation. Consequently, differentiation with respect to $\bm{\xi}$ acts only on the trunk networks, as shown in Eq. \eqref{eq:spinet_grad}.
Each trunk network
\(
\mathrm{tr}^j:\mathbb{R}\rightarrow\mathbb{R}^{p\times r}
\)
takes a scalar coordinate $\xi_j$ as input and outputs a vector of size $pr$, which is reshaped to size $p\times r$. Let $n_j$ denote the number of sampled points along the $j$-th coordinate direction, and let $C_f$ denote the computational cost of a single forward-mode derivative evaluation of a trunk network.
In the separable setting, forward-mode automatic differentiation evaluates coordinate derivatives independently for each trunk network. Since each trunk network receives a scalar input, the effective input dimension for differentiation is one, placing the residual evaluation firmly in the regime where FAD is computationally favorable. Assuming residuals are evaluated independently at each discretization point and that no cross-point derivative reuse is available, the cost of forward-mode differentiation scales as follows:
\begin{equation}
C_{\mathrm{sep}}
\propto
\sum_{j=1}^{d} n_j \, C_f .
\end{equation}
By contrast, a non-separable architecture that maps the full coordinate vector
$\bm{\xi}\in\mathbb{R}^d$ to the solution field requires forward-mode differentiation at every point on the tensor-product grid. The corresponding cost scales as,
\begin{equation}
C_{\mathrm{non\text{-}sep}}
\propto
\left(\prod_{j=1}^{d} n_j\right) C_f .
\end{equation}
After computing the coordinate-wise derivatives, these terms are combined through outer-product constructions and contracted with the branch coefficients to form the final operator output. This combination step is derivative-free and is required, in some form, whether a separable or non-separable architecture is employed. This cost is not included in the forward-mode differentiation cost $C_{\mathrm{sep}}$, which accounts exclusively for derivative evaluations.

In summary, the advantage of forward-mode automatic differentiation in SPINONet arises from two structural properties: first, the scaling of FAD with the input dimension, and second, the confinement of all coordinate dependence to separable, one-dimensional trunk networks. Together, these properties significantly reduce the cost of computing the residual-based loss function. As a consequence of reduced computational load, lower memory requirements, and the avoidance of derivative evaluations over the whole unrolled spatio-temporal grid, it is reasonable to expect that the combination of FAD and separable trunk architectures supports an energy-efficient training paradigm for SPINONet. While the precise energy savings depend on implementation details at both the software and hardware levels, the analysis presented here provides a system-agnostic surrogate that captures the fundamental computational advantages of this design.
Algorithm \ref{alg:spinet_fad} summarizes forward-mode residual evaluation in SPINONet for a single collocation point. The pointwise formulation is adopted for notational clarity; in practice, trunk evaluations and forward-mode derivative propagation are executed simultaneously over all collocation points using appropriate tensor operations.
\begin{algorithm}[ht!]
\caption{Forward-Mode Residual Evaluation in SPINONet}
\label{alg:spinet_fad}
\begin{algorithmic}[1]
\REQUIRE Trunk networks $\{\mathrm{tr}^j\}_{j=1}^{d}$; 
set of differentiated coordinates $\mathcal D \subseteq \{1,\dots,d\}$; 
branch coefficients $\mathcal B(\bm u)$; collocation point $\bm c$.
\FOR{each coordinate direction $j=1,\dots,d$}
    \STATE Evaluate trunk features $\mathrm{tr}^j(\xi_j)$ at $\bm c$.
    \IF{$j \in \mathcal D$}
        \STATE Compute forward-mode derivative $\partial \mathrm{tr}^j / \partial \xi_j$.
    \ENDIF
\ENDFOR
\STATE Assemble required trunk derivatives via separable outer products.
\STATE Contract with $\mathcal B(\bm u)$ to obtain $\mathcal G_\theta(\bm u)$ and required coordinate derivatives.
\STATE Evaluate PDE residual.
\end{algorithmic}
\textbf{Output}: Physics-informed residual at $\bm c$.
\end{algorithm}
\subsection{Surrogate Gradient Training in Branch Network}
\label{subsec:surrogate}
The spike-generation mechanism in VSNs is governed by a discontinuous heaviside function, which precludes the direct application of standard backpropagation. In particular, derivatives of the form $\partial \widetilde y_{\tau} / \partial m_{\tau}$ are undefined in the classical sense. As a result, standard gradient-based optimization methods cannot be directly applied to networks containing spiking neurons. To enable end-to-end training, we employ the surrogate gradient method \cite{neftci2019surrogate,yamazaki2022spiking}, in which the discontinuous spike function is replaced by a smooth approximation during the backward pass, while the original hard thresholding operation is retained in the forward pass. This approach preserves the event-driven dynamics of spiking neurons during inference, while providing meaningful gradients for parameter updates during training.

%
Applying the chain rule through spike time steps in VSN dynamics, the required derivatives can be expressed in terms of the membrane potential and spike variables as
\begin{equation}
\begin{aligned}
\dfrac{\mathrm d y^{(i)}}{\mathrm d z^{(j)}} &=
\dfrac{\partial y^{(i)}}{\partial \widetilde y^{(i)}}
\dfrac{\partial \widetilde y^{(i)}}{\partial m^{(i)}}
\prod_{k=j}^{i-1}
\left(
\dfrac{\partial m^{(k+1)}}{\partial m^{(k)}}
+
\dfrac{\partial m^{(k+1)}}{\partial \widetilde y^{(k)}}
\dfrac{\partial \widetilde y^{(k)}}{\partial m^{(k)}}
\right)
\dfrac{\partial m^{(j)}}{\partial z^{(j)}},
\quad i < j,
\\
\dfrac{\mathrm d y^{(i)}}{\mathrm d z^{(j)}} &=
\dfrac{\partial y^{(i)}}{\partial \widetilde y^{(i)}}
\dfrac{\partial \widetilde y^{(i)}}{\partial m^{(i)}}
\dfrac{\partial m^{(i)}}{\partial z^{(i)}}
+
\dfrac{\partial y^{(i)}}{\partial z^{(i)}},
\quad i = j.
\end{aligned}
\label{equation: grads_dy_dz}
\end{equation}

The main challenge in evaluating the above expressions lies in computing
$\partial \widetilde y^{(\tau)} / \partial m^{(\tau)}$, since spike generation is governed by a discontinuous threshold function. Following standard surrogate-gradient approaches, we retain the hard threshold in the forward pass,
\begin{equation}
\widetilde y_{\tau} = \mathbb H\!\left(m_{\tau} - \mathcal T_h\right),
\end{equation}
and approximate its derivative during backpropagation by a smooth function as,
\begin{equation}
\frac{\partial \widetilde y^{(\tau)}}{\partial m^{(\tau)}}=\frac{\partial \mathbb H\!\left(m_{\tau} - \mathcal T_h\right)}{\partial m^{(\tau)}}
\;\widehat{=}\;
\frac{1}{1 + K_s \left| m^{(\tau)} - \mathcal T_h \right|},
\label{equation: surrogate_grad}
\end{equation}
where $K_s$ is a slope parameter controlling the sharpness of the surrogate gradient. A different surrogate gradient can also be applied depending on the needs of the task at hand. 

This surrogate-gradient formulation enables stable gradient-based optimization of the VSN-based branch network in SPINONet while preserving the discontinuous spiking behavior during inference. It should be noted that because surrogate gradients only approximate the true derivative, their use introduces an approximation error in gradient computation. Incorporating VSNs into the coordinate-dependent trunk networks would therefore contaminate the computation of physics-informed residuals, leading to biased PDE learning. For this reason, VSNs are employed exclusively in the branch network, where they affect only the generation of coefficient vectors and do not interfere with spatio-temporal differentiation.
Algorithm~\ref{alg:spinet_surr} summarizes the surrogate-gradient backpropagation procedure used to compute gradients for the VSN-based branch network. For clarity, the procedure is presented for a single input sample; in practice, it is applied in minibatches.
\begin{algorithm}[ht!]
\caption{Surrogate Gradient Backpropagation in the SPINONet Branch Network}
\label{alg:spinet_surr}
\begin{algorithmic}[1]
\REQUIRE Input $\bm u$; branch parameters $\theta_B$; loss $\mathcal L$.
\STATE Perform forward pass through VSN using 
$\widetilde y^{(\tau)}=\mathbb H(m^{(\tau)}-\mathcal T_h)$ to obtain $\mathcal B(\bm u)$.
\STATE Evaluate loss $\mathcal L$ using physics-informed and data supervision terms.
\STATE Backpropagate through VSN dynamics using surrogate derivatives 
$\partial \widetilde y^{(\tau)}/\partial m^{(\tau)}$.
\STATE Accumulate gradients $\nabla_{\theta_B}\mathcal L$ for branch parameters.\\
\end{algorithmic}
\textbf{Output}: Gradients $\nabla_{\theta_B}\mathcal L$ for updating $\theta_B$.

\end{algorithm}

In practice, VSNs can be unrolled along the spike-time dimension and accept inputs represented as spike trains of appropriate length. However, for regression-based operator learning, we observe that directly supplying continuous-valued inputs, repeated over all $T_s$ spike time steps, yields the best empirical performance. This strategy avoids unnecessary encoding overhead while retaining sparse, event-driven computation.

\section{Numerical Examples}
\label{section:results}
This section presents numerical studies demonstrating the performance of SPINONet on representative PDE benchmarks. We consider three problems of increasing complexity, namely, (i) the viscous Burgers equation, involving one spatial and one temporal dimension, (ii) heat equation with parametric diffusion, involving two spatial dimensions, one temporal dimension and one parameteric dependency, and, (iii) Eikonal equation with parametric boundaries, involving two spatial dimensions. These examples collectively test nonlinear dynamics, parametric dependence, and geometric input representations. In all cases, we learn a solution operator $\mathcal{G}:\mathcal{U}\rightarrow\mathcal{Y}$ that maps admissible inputs $\bm u \in \mathcal{U}$ (e.g., initial conditions, parameters, or boundary descriptions) to spatio-temporal solution fields $\bm y=\mathcal{G}(\bm u) \in \mathcal{Y}$. Model predictions are evaluated on unseen test samples and queried on structured grids. We compare SPINONet against a vanilla separable physics-informed DeepONet, hereafter referred to as \textit{vanilla baseline} and, where feasible, a Physics-informed DeepONet (PI-DeepONet), assessing both predictive accuracy and computational efficiency.

To improve readability, architectural choices, discretization strategies, and training configurations for all examples are summarized in Table~\ref{tab:setup_summary}. Table \ref{tab: combined} presents the errors observed in testing data and compares SPINONet against the PI-DeepONet architecture as well. The reported relative $L^2$ errors are computed as,
\begin{equation}
\mathrm{Rel}\text{-}L^2
=
\frac{1}{N_{\mathrm{test}}}
\sum_{i=1}^{N_{\mathrm{test}}}
\frac{
\left\|
\widehat{\mathbf{s}}^{(i)} - \mathbf{s}^{(i)}
\right\|_2
}{
\left\|
\mathbf{s}^{(i)}
\right\|_2
},
\end{equation}
where $\mathbf{s}^{(i)}$ denotes the reference solution field for the $i$-th test sample, $\widehat{\mathbf{s}}^{(i)}$ denotes the corresponding predicted solution field, and $N_{\mathrm{test}}$ is the total number of test samples. The norm $\|\cdot\|_2$ is taken over all spatio-temporal grid points.
While the errors observed are in a similar range, the training time per epoch is significantly faster in SPINONet, supported by its separable architecture and use of forward mode automatic differentiation. Below, we focus on qualitative behavior, accuracy, sparsity, and scaling characteristics.

\begin{table}[ht!]
\centering
\caption{Summary of architectures, discretization, and training configurations for all numerical examples.}
\label{tab:setup_summary}
\begin{tabular}{lccc}
\toprule
 & Burgers & Heat (Parametric) & Eikonal \\
\midrule
Spatial dimension & 1D & 2D & 2D \\
Temporal dimension & 1 & 1 & -- \\
Parameter & IC & $\alpha$ and IC & Geometry \\
Total dim., $d$ & 2 & 4 & 2 \\
\midrule
Branch hidden layers & $6 \times 100$ & $5 \times 50$ & $5 \times 50$ \\
Branch input size & 101 & 1 & 400\\
Trunk hidden layers & $6 \times 50$ & $5 \times 50$ & $5 \times 50$ \\
Separable ranks $(p,r)$ & 20 & 50 & 25 \\
\midrule
Test samples & 1000 & 150 & 1000 \\
Collocation points & 2,601 & 923,521 & 1,600 \\
\midrule
Optimizer & Adam & Adam & Adam \\
Epochs & 40k & 100k & 40k

\\
\bottomrule
\end{tabular}
\end{table}

\begin{table}[ht!]
\centering
\caption{Comparison of test $L^2$ error and runtime per epoch for PI-DeepONet, Vanilla Baseline, and SPINONet across Burgers, Heat, and Eikonal equations. Average spiking activity averaged across all VSN layers of the branch net of SPINONet is also shown.}
\label{tab: combined}
\begin{tabular}{ccccc}
\toprule
Equation & Model & $L^2$ error & Runtime (s)/Epoch & Spiking Activity \\
\midrule

\multirow{3}{*}{Burgers}
& PI-DeepONet & 0.05 & $\sim$ 0.604 & - \\
& Vanilla Baseline & 0.06 & $\sim$ 0.012 & - \\
& SPINONet & 0.07 & $\sim$ 0.013 & 53.41 \\
\midrule

\multirow{3}{*}{Heat}
& PI-DeepONet & - & - & - \\
& Vanilla Baseline & 0.10 & $\sim$ 0.097 & - \\
& SPINONet & 0.09 & $\sim$ 0.095 & 46.33 \\
\midrule

\multirow{3}{*}{Eikonal}
& PI-DeepONet & 0.005 & $\sim$ 0.261 & - \\
& Vanilla Baseline & 0.007 & $\sim$ 0.039 & - \\
& SPINONet & 0.016 & $\sim$ 0.041 & 32.22 \\
\bottomrule
\end{tabular}
\end{table}

\begin{figure}[ht!]
    \centering
    \includegraphics[width=0.75\textwidth]{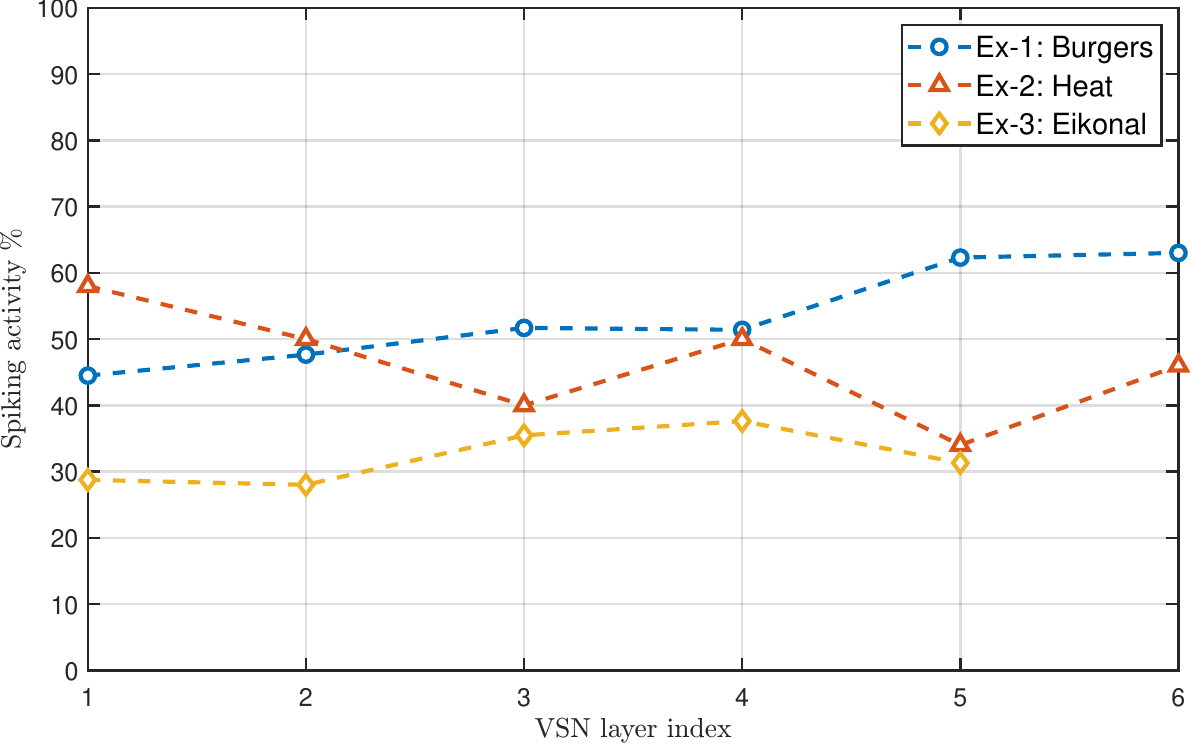}
    \caption{Mean per-layer spiking activity in the branch-network VSN layers across all numerical examples.}
    \label{fig:spiking_activity_all_examples}
\end{figure}



\subsection{Example E-I: Viscous Burgers Equation}

We first consider the one-dimensional viscous Burgers equation, a canonical nonlinear convection--diffusion model that exhibits wave steepening and viscous regularization:
\begin{equation}
\frac{\partial u}{\partial t}(x,t)
+
u(x,t)\frac{\partial u}{\partial x}(x,t)
=
\nu \frac{\partial^2 u}{\partial x^2}(x,t),
\qquad (x,t)\in\Omega\times\mathcal{T}.
\label{eq:burgers_pde}
\end{equation}
Periodic boundary conditions are imposed, and initial conditions are sampled from a Gaussian process. The learning objective is to approximate the operator mapping $u(x,0)\mapsto u(x,t)$. To test the performance of the trained model, predictions were made on a spatio-temporal grid of size 101$\times$ 101.

\begin{figure}[ht!]
    \centering

    \begin{subfigure}[t]{0.9\textwidth}
        \centering
        \includegraphics[width=0.85\textwidth]{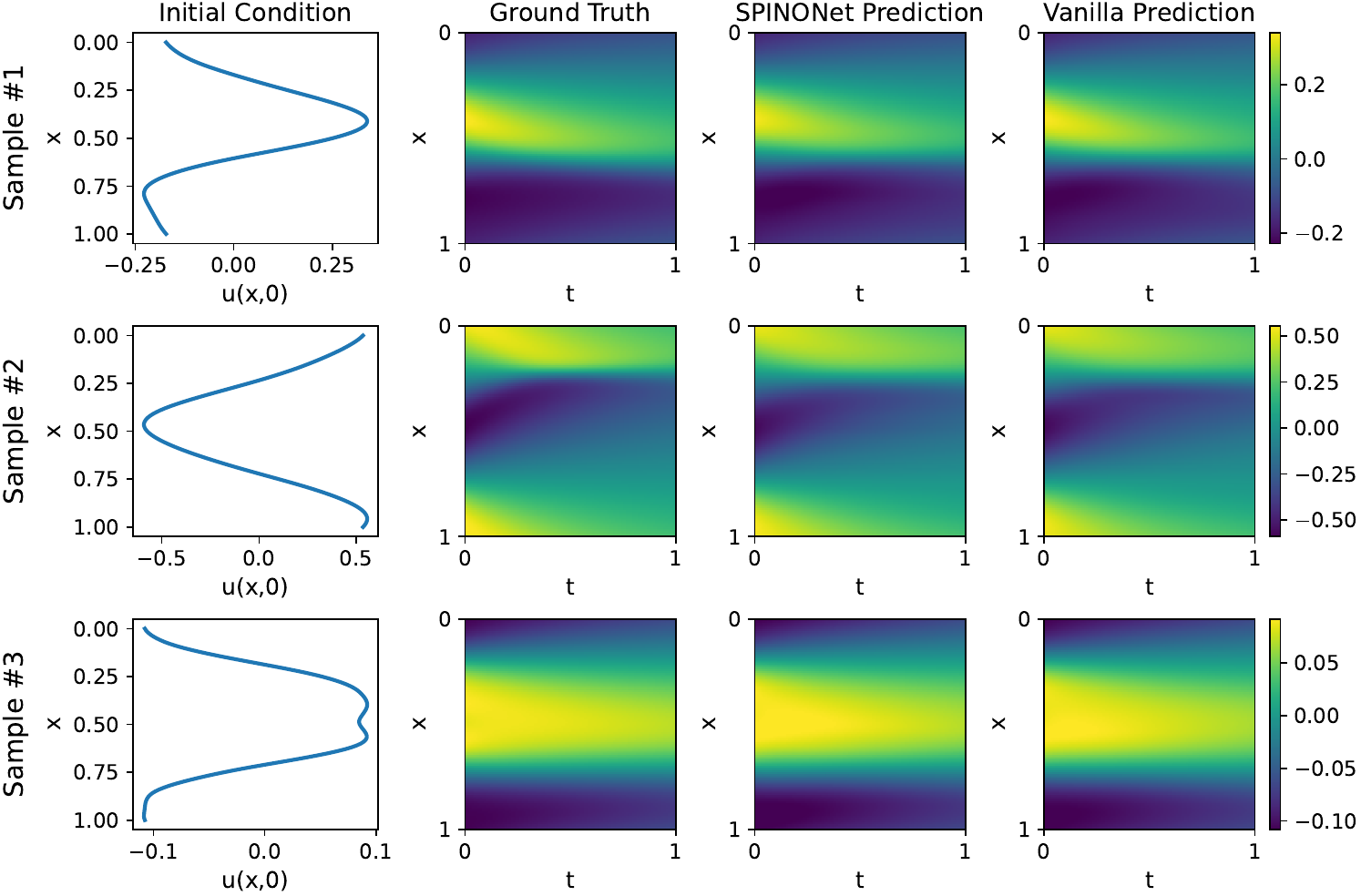}
        \caption{Initial conditions, ground-truth solutions, and corresponding predictions produced by SPINONet and the vanilla baseline for representative test samples.}
        \label{fig:burgers_2d_pred}
    \end{subfigure}

    \vspace{0.5em}

    \begin{subfigure}[t]{0.45\textwidth}
        \centering
        \includegraphics[width=\textwidth]{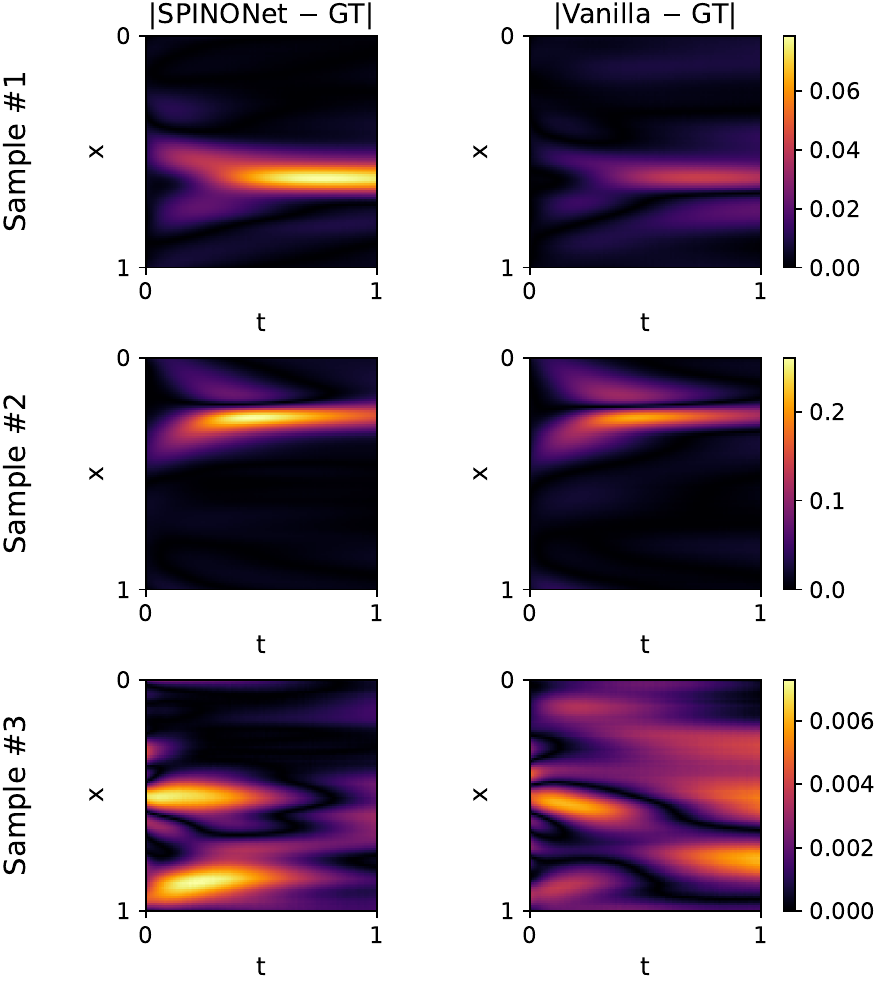}
        \caption{Absolute error fields corresponding to the predictions shown above for SPINONet and the vanilla baseline.}
        \label{fig:burgers_2d_err}
    \end{subfigure}

    \caption{Viscous Burgers equation: Representative comparisons between SPINONet and the vanilla baseline on unseen test samples. Rows correspond to different test cases, while columns show solution fields and error distributions.}
    \label{fig:burgers_2d_comparison}
\end{figure}

Figure~\ref{fig:burgers_2d_comparison} shows representative spatio-temporal predictions on unseen test samples. SPINONet accurately captures the evolution of the solution and closely matches the reference fields. The error patterns are comparable to those of the vanilla baseline, with discrepancies primarily localized near regions of steep gradients.
Across test samples, SPINONet exhibits sparse activity in the branch network, with mean firing rates between approximately $45\%$ and $65$\% across VSN layers (Fig.~\ref{fig:spiking_activity_all_examples}). The average relative $L^2$ error is $0.07$, compared to $0.06$ for the vanilla baseline and $0.05$ for PI-DeepONet.

\begin{figure}[ht!]
    \centering
    
    \begin{subfigure}[t]{0.49\textwidth}
        \centering
        \includegraphics[width=\textwidth]{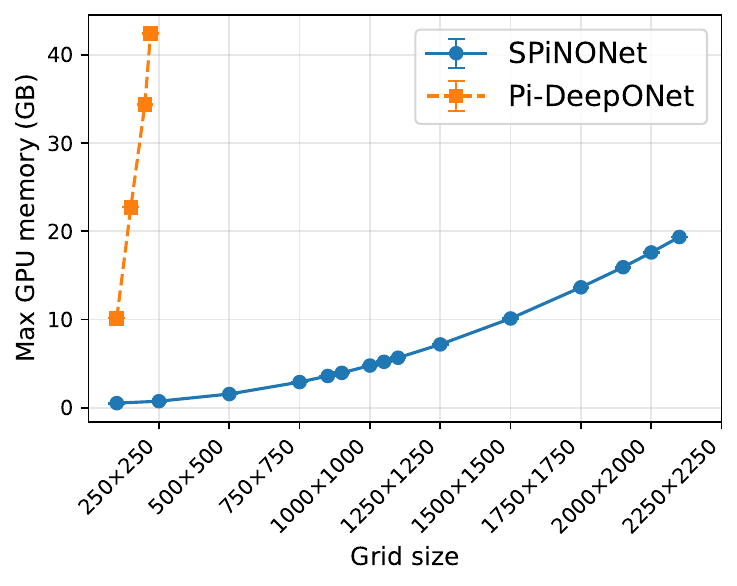}
        \caption{GPU memory consumption.}
        \label{fig:burgers_memory_comparison}
    \end{subfigure}
    %
    %
    \begin{subfigure}[t]{0.49\textwidth}
        \centering
        \includegraphics[width=\textwidth]{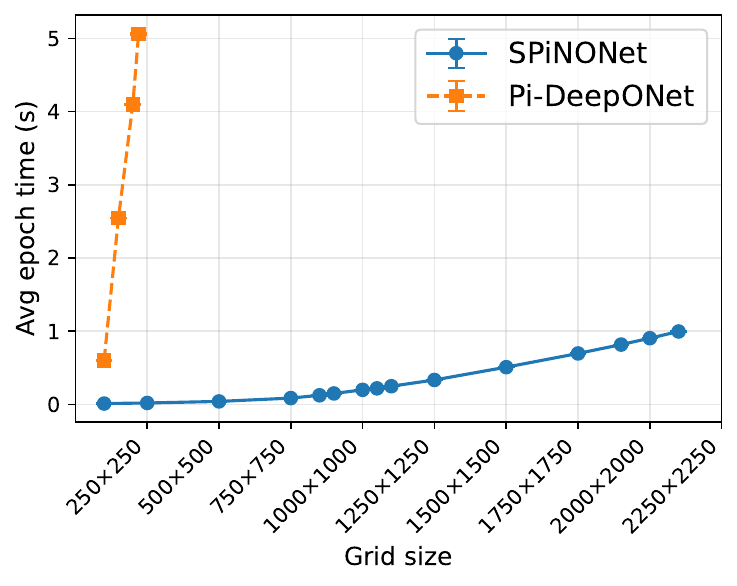}
        \caption{Average training time per epoch.}
        \label{fig:burgers_time_comparison}
    \end{subfigure}
    
    \caption{
    Viscous Burgers equation: Comparison of SPINONet and PI-DeepONet computational cost as a function of grid size. 
    The top panel shows GPU memory consumption, and the bottom panel shows average training time per epoch. Error bars indicate one standard deviation computed over independent runs.}
    
\end{figure}

Beyond accuracy, SPINONet demonstrates favorable computational scaling. Figures~\ref{fig:burgers_memory_comparison} and \ref{fig:burgers_time_comparison} compare GPU memory usage and average training time per epoch against PI-DeepONet across increasing grid resolutions. SPINONet consistently requires substantially less memory and exhibits slower growth in runtime as resolution increases.
These gains arise from the separable operator representation in SPINONet, which enables coordinate-wise trunk evaluation and algebraic assembly of the solution field. In contrast, PI-DeepONet evaluates the trunk network over the full spatio-temporal grid, leading to rapidly increasing computational and memory costs. This structural difference explains the superior scalability observed for SPINONet.


\subsection{Example E-II: Heat Equation with Parametric Diffusion}

We next consider the two-dimensional heat equation with a parametric diffusion coefficient:
\begin{equation}
\frac{\partial u}{\partial t}(x,y,t)
=
\alpha\left(
\frac{\partial^2 u}{\partial x^2}(x,y,t)
+
\frac{\partial^2 u}{\partial y^2}(x,y,t)
\right),
\qquad (x,y,t)\in\Omega\times\mathcal{T}.
\label{eq:heat_pde}
\end{equation}
Zero Dirichlet boundary conditions are imposed. The learned operator maps the scalar initial condition and diffusion parameter to the full spatio-temporal solution field,
$u(x,y,0;\alpha)\mapsto u(x,y,t;\alpha)$.
The diffusion parameter is incorporated into the trunk input as $\alpha^{1/2}$ to improve numerical conditioning. In this example, predictions from the trained model were made on a spatio-temporal grid of size 51$\times$51$\times$51, with the grid resolved for 150 values of $\alpha$.

\begin{figure}[ht!]
    \centering
    \includegraphics[width=\textwidth]{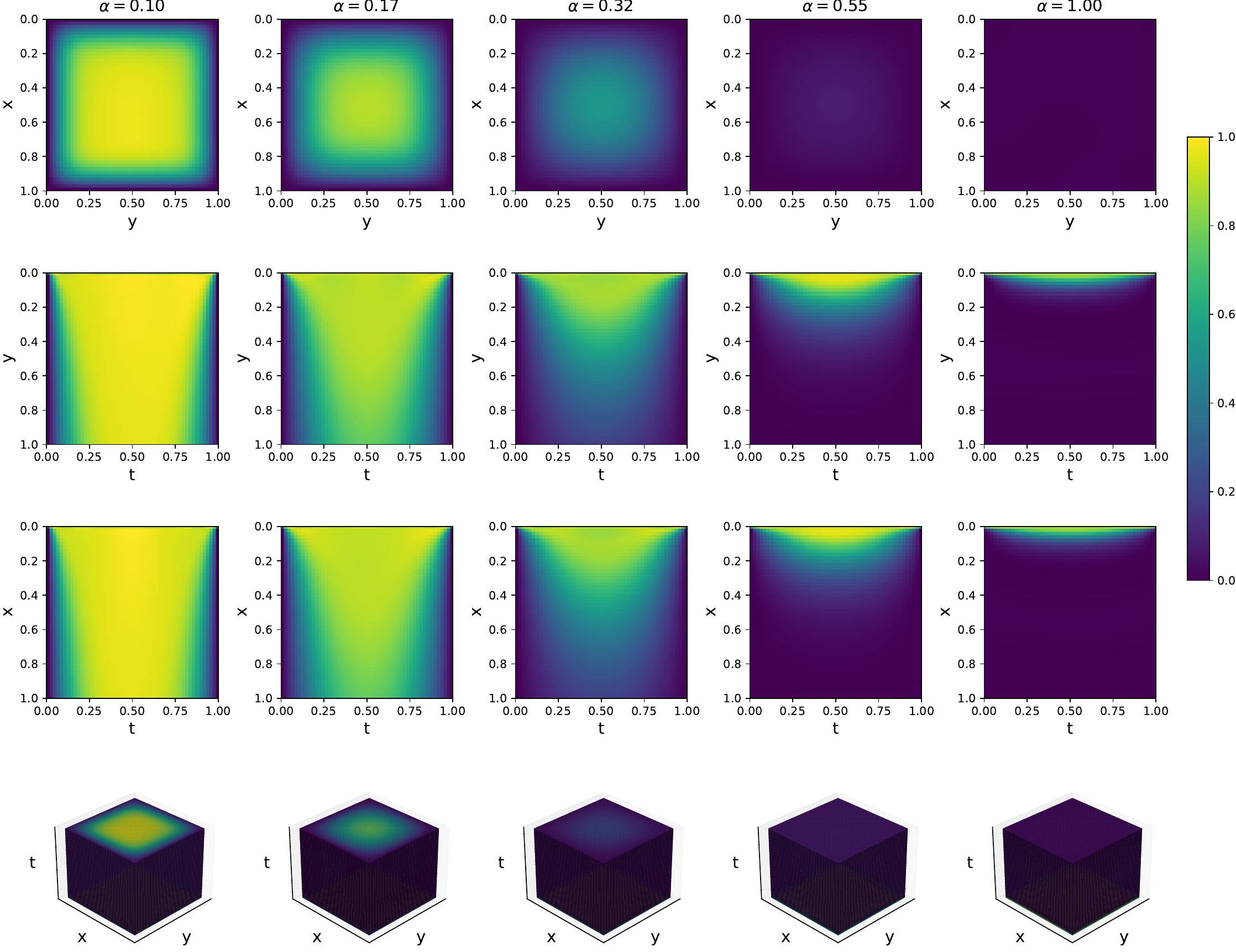}
    \caption{
    Heat equation with Parameteric Diffusion: SPINONet predictions at selected values of $\alpha$. Shown are representative planar slices and volumetric renderings of $u(x,y,t)$.}
    \label{fig:heat_4d_planes_voxels}
\end{figure}

\begin{figure}[ht!]
    \centering
    \includegraphics[width=\textwidth]{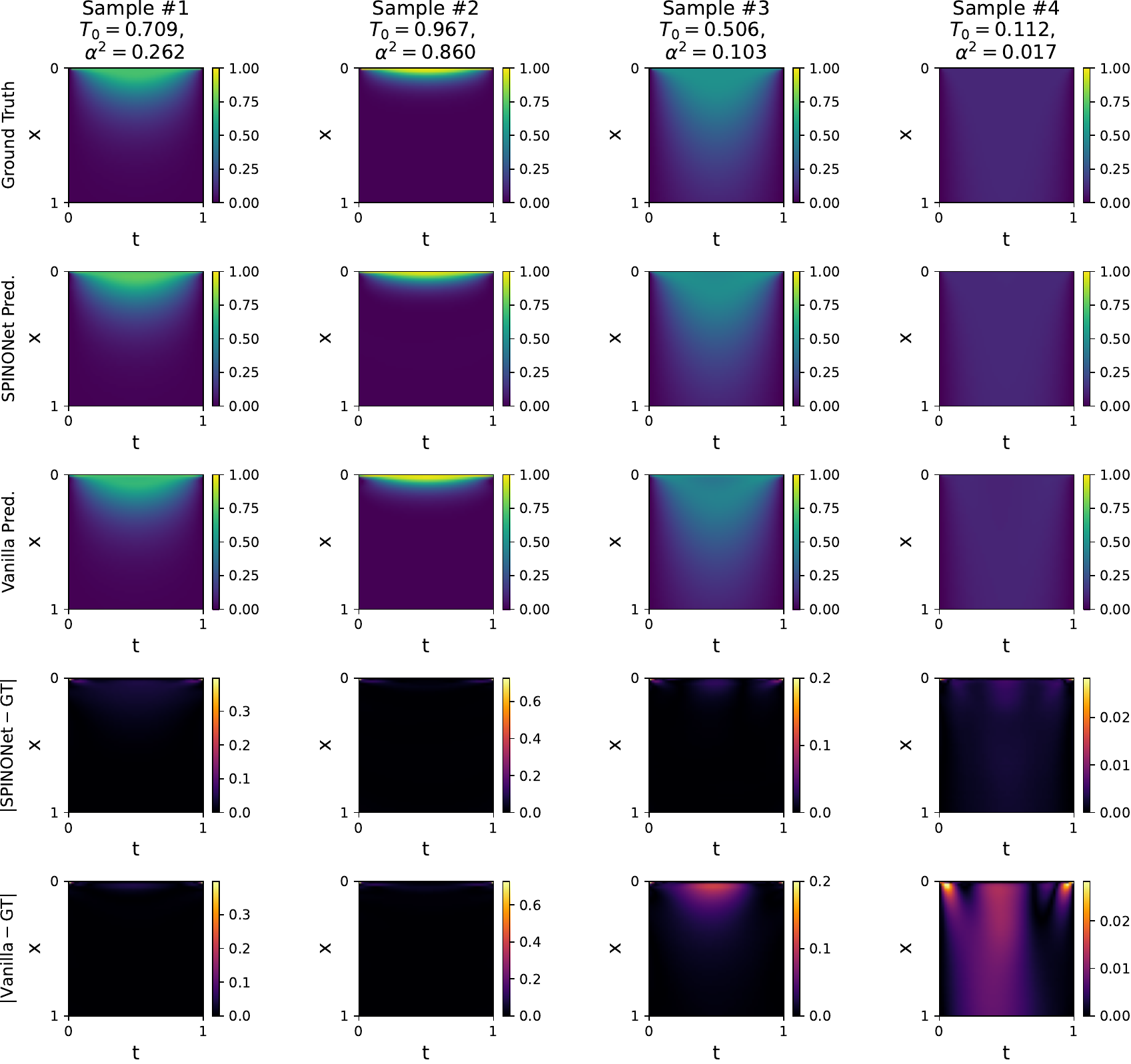}
    \caption{
    Heat equation with Parameteric Diffusion: Comparison of reference solutions, SPINONet predictions, vanilla baseline predictions, and absolute errors. Columns correspond to different test cases, while rows show solution fields and error
    distributions.}
    \label{fig:heat_4d_comparison}
\end{figure}

Figure~\ref{fig:heat_4d_planes_voxels} visualizes representative SPINONet predictions for different values of $\alpha$. The model reproduces the expected diffusive behavior, with larger diffusion coefficients producing faster temporal decay and increased spatial smoothing.
Figure~\ref{fig:heat_4d_comparison} compares SPINONet and vanilla baseline predictions against reference solutions. SPINONet achieves a mean relative $L^2$ error of $0.09$, slightly outperforming the vanilla baseline ($0.10$), while maintaining sparse branch computation with firing rates between approximately $34\%$ and $58$\%. PI-DeepONet failed to converge for this example due to prohibitive memory requirements. These results further highlight the scalability advantages of the separable formulation in SPINONet. Training SPINONet using data loss alone yields significantly larger errors, underscoring the importance of physics-informed constraints for this problem.

\subsection{Example E-III: Eikonal Equation with Parametric Boundaries}

Finally, we consider a two-dimensional Eikonal equation to assess SPINONet’s ability to handle geometric input representations:
\begin{equation}
\sqrt{s_x(x,y)^2 + s_y(x,y)^2} = 1,
\qquad (x,y)\in\Omega,
\label{eq:eikonal_pde_shorthand}
\end{equation}
with Dirichlet boundary conditions $s=0$ on $\partial\Omega$. The solution corresponds to the signed distance function to the boundary. The input to the model is a discretized representation of a circular boundary, and the output is the signed distance field over the spatial domain. Both SPINONet and the vanilla baseline employ separable trunk representations with $\bm{\xi}=(x,y)$.
During training, we observed that purely physics-informed formulations may converge to degenerate solutions. To mitigate this, a small supervised loss is added on the residual grid. 200 training samples were used for this purpose. Predictions using the trained network were made on a spatial grid of size $200\times200$.

\begin{figure}[ht!]
    \centering
    \includegraphics[width=0.98\textwidth]{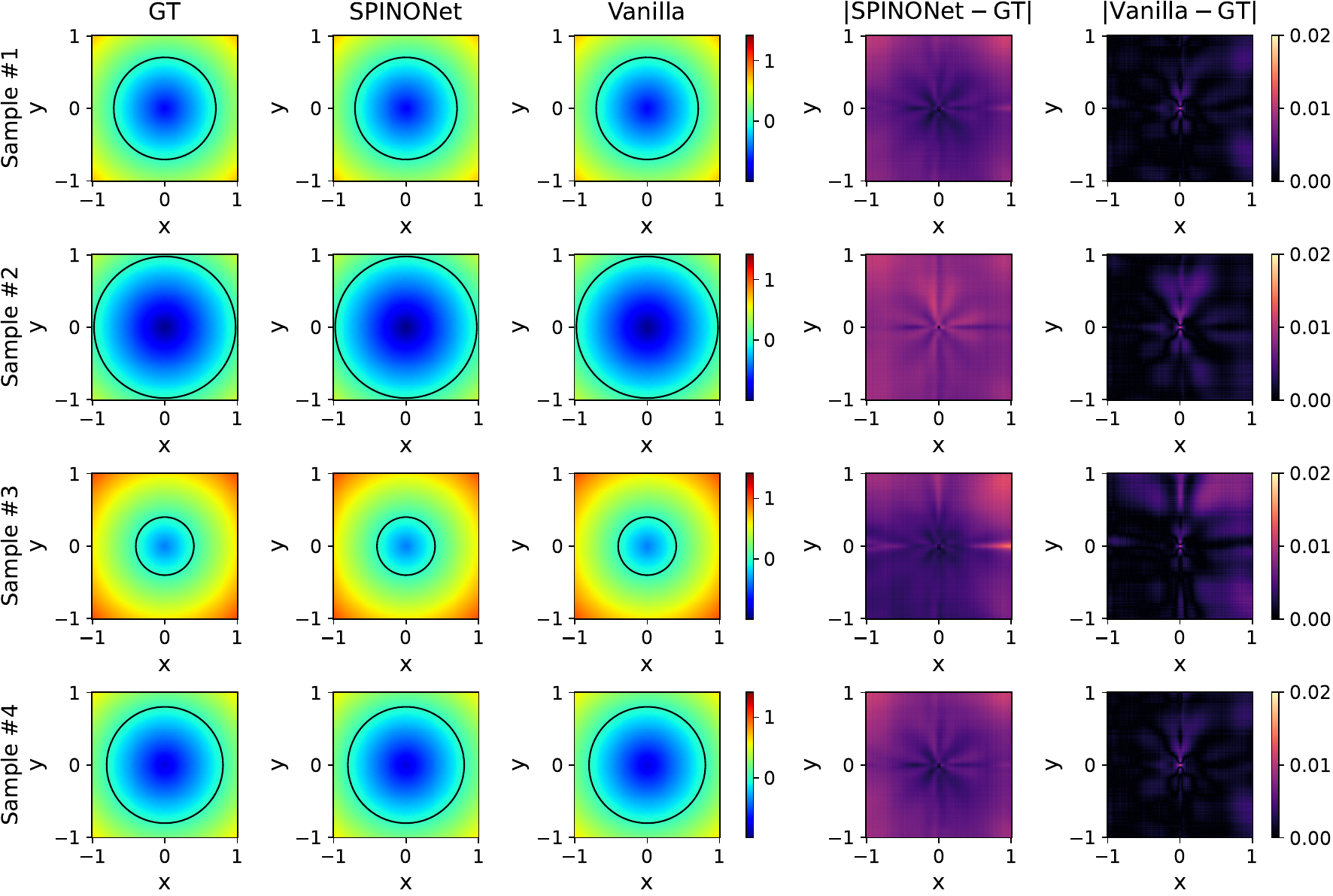}
    \caption{
    Eikonal equation with parametric boundaries: Comparison of reference solutions, SPINONet predictions,
    vanilla baseline predictions, and corresponding absolute errors. Rows correspond to test samples; columns show solution fields and absolute errors.}
    \label{fig:eikonal_2d_comparison}
\end{figure}

Figure~\ref{fig:eikonal_2d_comparison} shows representative predictions on test samples. Both models accurately recover the signed distance fields, with errors concentrated near the boundary.
SPINONet achieves a relative $L^2$ error of $0.016$, compared to $0.007$ for the vanilla baseline and $0.005$ for PI-DeepONet, while maintaining sparse branch activity with firing rates between approximately $28\%$ and $37\%$. Runtime per epoch remains comparable to the vanilla baseline, whereas PI-DeepONet incurs substantially higher computational cost.

\section{Conclusion}\label{section:conclusion}
This work introduced the Separable Physics-informed Neuroscience-inspired Operator Network (SPINONet), a physics-informed operator-learning framework designed to reduce computational cost in settings where learned operators are evaluated repeatedly, without sacrificing accuracy or compatibility with residual-based training. The key idea behind SPINONet is to exploit the separable structure of physics-informed operator networks to enable sparse, event-driven computation in the branch pathway using neuroscience-inspired spiking neurons, while preserving continuous, coordinate-differentiable trunk networks required for computing spatio-temporal derivatives. Specifically, we employ Variable Spiking Neurons (VSNs) within the SPINONet architecture due to their graded, continuous-valued spike formulation and demonstrated effectiveness in regression and operator-learning settings. This separation allows sparsity to be introduced without altering the physics-informed loss formulation and directly targets redundant computation during repeated operator queries on grids.

We evaluated SPINONet on multiple PDEs with coupled spatial, temporal, and parametric dependencies. Collectively, these benchmarks require learning operators over high-dimensional input spaces, providing a nontrivial test of scalability. For instance, the parameterized heat equation involves two spatial dimensions, one temporal dimension, and one parameter dimension, resulting in a four-dimensional operator-learning problem.
%
Across all examples, SPINONet consistently produced accurate solution fields while exhibiting clear gains in computational efficiency and scalability compared to dense operator-learning baselines. In particular, SPINONet achieved competitive prediction accuracy while maintaining substantial sparsity in the branch network, with firing rates ranging from approximately 28\% to 65\% across problems. Despite this sparsity, the model remained fully compatible with physics-informed residual training, as the continuous and differentiable trunk pathways required for computing spatio-temporal derivatives were preserved throughout. Runtime per epoch remained comparable to the dense separable baseline, while non-separable architectures such as PI-DeepONet incurred significantly higher computational and memory costs and, in some cases, failed to converge.

The benefits of the proposed architecture were especially apparent in challenging regimes. In the parametric heat-equation example, SPINONet remained tractable, where PI-DeepONet became prohibitively memory-intensive. In the Eikonal example, the addition of a small supervised loss successfully mitigated degenerate solutions encountered under purely physics-informed training, improving stability without changing the residual formulation. Across all problems, SPINONet demonstrated smooth scaling with increasing spatial and temporal resolution and produced stable predictions across a wide range of evaluation grids.

Overall, these results show that sparse, event-driven computation can be integrated into separable physics-informed operator learning in a principled and practical way. While challenges common to deep learning remain, such as sensitivity to optimization settings and initialization, the proposed framework provides a promising step toward energy- and resource-aware operator learning. Future work will explore hardware-aware implementations, sparsity-aware training objectives, and extensions to higher-dimensional and multi-physics systems. Taken together, SPINONet offers a compelling approach for accurate and scalable operator evaluation under realistic computational and power constraints.

\section*{Acknowledgment}
The first author acknowledges financial support from the Ministry of Education, India, through the Prime Minister's Research Fellows (PMRF) scholarship. The third author
acknowledges the support from the U.S. Department of Energy (DOE), Office of Science, Office of Advanced Scientific Computing Research, under Award No. DE-SC0024162. The fourth author acknowledges the financial support received from the Anusandhan National Research Foundation (ANRF) via grant no. CRG/2023/007667 and from the Ministry of Port and Shipping via letter no. ST-14011/74/MT (356529).

\appendix
\section{Degenerate convergence modes in physics-informed training for the Eikonal equation}
\label{appendix:degenerate_eikonal}

In the Eikonal example, we observed that purely physics-informed training can sometimes converge to solutions that satisfy the PDE constraint numerically but fail to represent the correct signed distance field. This behavior is not unique to our model and was observed for both the vanilla baseline and SPINONet during early experiments. The core issue is that the Eikonal constraint $\|\nabla s\|=1$ does not, by itself, uniquely select the correct signed distance field unless additional information anchors the sign structure and the boundary condition behavior in a consistent way across the domain. As a result, optimization may settle into alternative solutions that appear reasonable under the residual but do not correspond to the desired interior-exterior signed distance map.

\begin{figure}[ht!]
    \centering
    \includegraphics[width=0.98\textwidth]{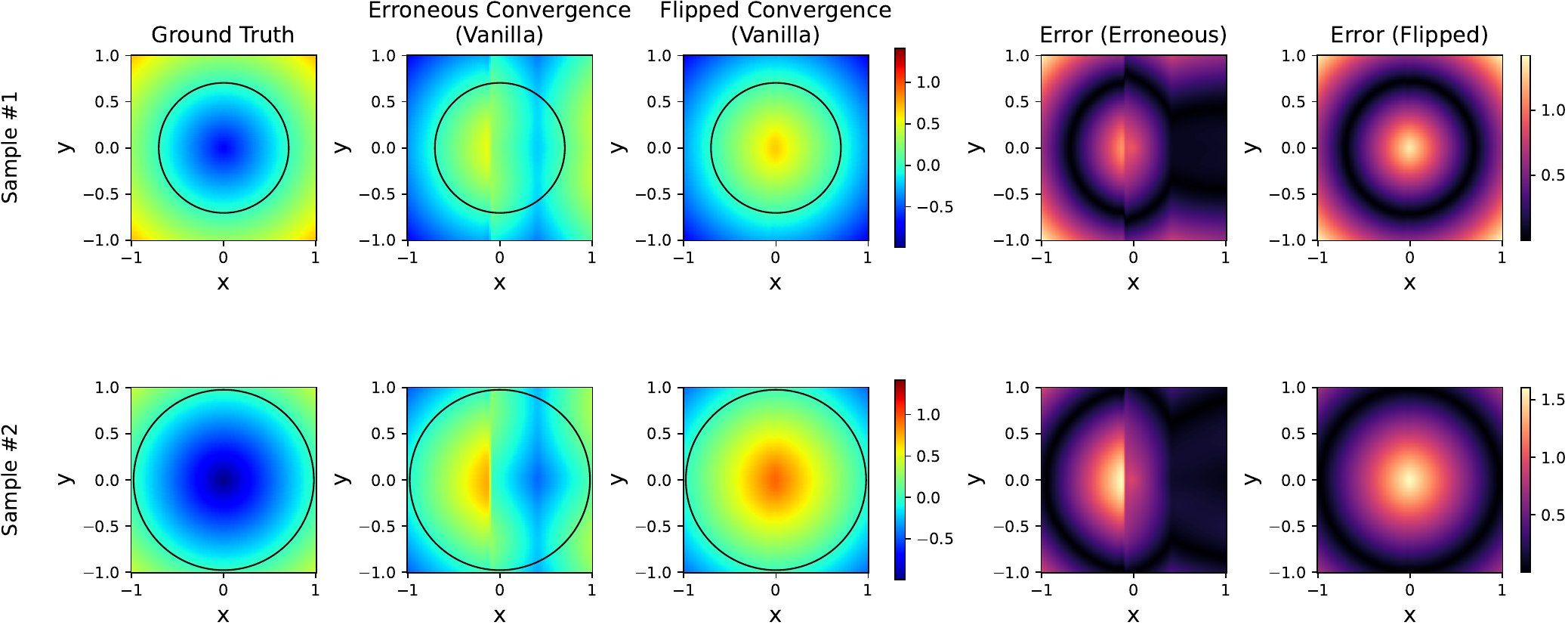}
    \caption{
    Eikonal equation with parametric boundaries: Representative degenerate convergence modes for the Eikonal equation using the vanilla baseline. Erroneous convergence and flipped-sign convergence, along with the corresponding absolute errors.
    }
    \label{fig:eikonal_vanilla_erroneous_vs_flipped}
\end{figure}
This behavior is illustrated in Fig.~\ref{fig:eikonal_vanilla_erroneous_vs_flipped}, where the vanilla baseline is shown to converge to two different failure modes depending on initialization and training dynamics: an \emph{erroneous convergence} case and a \emph{flipped convergence} case. In the erroneous case, the predicted field develops non-physical directional artifacts and vertical band-like structures, even though the boundary is still visually captured. In the flipped case, the network converges to a solution with the opposite sign convention, producing a field that resembles a valid distance map but assigns positive values outside and negative values inside. The corresponding error plots show that the erroneous solution produces large structured errors across the domain, while the flipped solution produces a characteristic ring-shaped error concentrated around the interface, reflecting the global sign reversal.

\begin{figure}[ht!]
    \centering
    \includegraphics[width=0.98\textwidth]{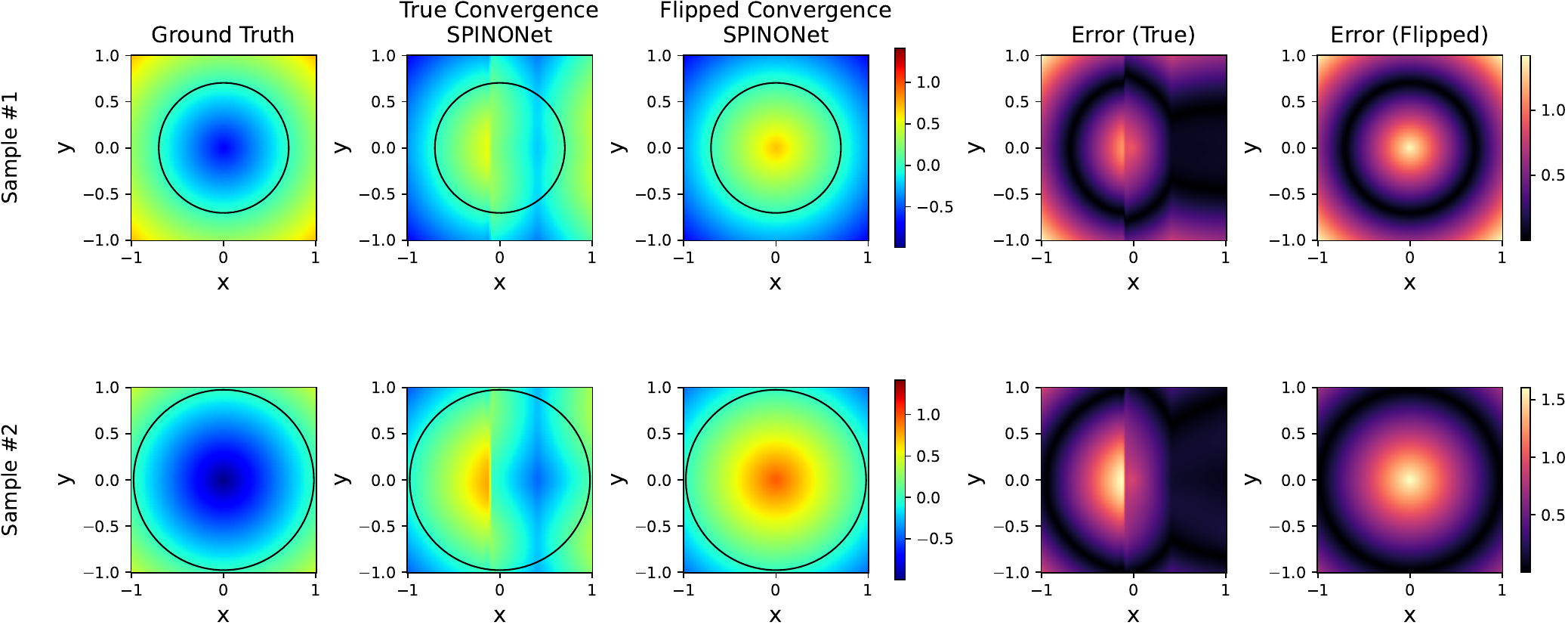}
    \caption{
    Eikonal equation with parametric boundaries: Comparison of true convergence and flipped-sign convergence for the Eikonal equation, with corresponding absolute errors, when using the proposed SPINONet.
    }
    \label{fig:eikonal_true_vs_flipped}
\end{figure}
Similar trends were observed when using SPINONet. Fig.~\ref{fig:eikonal_true_vs_flipped} compares a true convergence case against the flipped convergence behavior for the same PDE setup, when using SPINONet for prediction. In the true convergence case, the predicted signed distance field aligns closely with the ground truth, and the error remains small throughout the domain, with only faint star-like numerical patterns that are negligible in magnitude. In contrast, the flipped convergence again produces a smooth-looking field with the correct boundary location but the wrong sign assignment, leading to a large, spatially coherent error despite an apparently reasonable shape. This highlights that visual smoothness alone is not sufficient to guarantee correct convergence for Eikonal-based training, and that additional supervision or sign-consistent constraints are needed to prevent this degenerate behavior. As discussed in the respective example of the Numerical Illustration section, motivated by these observations, we augment the physics-only objective with a small supervised loss on the same grid used for residual enforcement. This additional signal acts as an anchor that discourages flipped or distorted solutions and consistently guides training toward the correct signed distance field across the full range of boundary instances.

\end{document}